\documentclass[a4paper,twocolumn,11pt]{quantumarticle}
\pdfoutput=1
\usepackage[utf8]{inputenc}
\usepackage{amsmath}
\usepackage[english]{babel}
\usepackage[T1]{fontenc}
\usepackage{amsthm}
\usepackage{amsfonts}
\usepackage{graphicx}
\usepackage{subcaption}
\usepackage{pgfplotstable}
\usepackage{nameref}
\usepackage{braket}
\usepackage{pgfplots}
\pgfplotsset{compat=1.18} 
\usepackage{filecontents}
\usepackage[table]{xcolor}
\usepackage{txfonts}
\usepackage{algorithm}
\usepackage{adjustbox}
\usepackage[noend]{algpseudocode}
\usepackage{float} 
\usepackage{hyperref}
\usepackage{cleveref}

\usepackage{cancel}
\usepackage{array}
\usepackage{lipsum}
\usepackage{siunitx}
\usepackage{booktabs}
\usepackage{pythontex}
\usepackage{tabularx}
\usepackage{longtable}
\usepackage{geometry}
\usepackage{hyphenat}
\usepackage{enumitem}
\usepackage{ulem}
\usepackage{setspace}
\usepackage{amsthm}
\usepackage{multirow}

\setlist[enumerate]{leftmargin=*,noitemsep,topsep=0pt}
\setlist[itemize]{leftmargin=*,noitemsep,topsep=0pt}

\usepackage{tikz}
\usetikzlibrary{arrows.meta,positioning,shapes.geometric,shapes.misc,calc,fit,backgrounds,shadows.blur}
\usetikzlibrary{shapes.geometric, arrows.meta, positioning}

\tikzstyle{block} = [rectangle, rounded corners, draw=black, fill=gray!10,
                     text width=6.5em, align=center, minimum height=3em, font=\small]
\tikzstyle{decision} = [diamond, draw=black, fill=gray!10,
                        text width=7em, aspect=2, align=center, inner sep=1pt, font=\small]
\tikzstyle{line} = [draw, -{Latex[length=2mm, width=1mm]}]
\tikzstyle{cloud} = [ellipse, draw=black, fill=blue!10, text width=8em, align=center, font=\small]

\begin{document}

\title{Emergency hub placement with a neutral-atom quantum computer}

\author{Sara Tarquini}
\affiliation{Gran Sasso Science Institute,  Viale Francesco Crispi 7, L'Aquila, Italy}
\email{sara.tarquini@gssi.it}
\orcid{}
\author{Matteo Vandelli}
\affiliation{Quantum Computing Solutions, Leonardo S.p.A., Via R. Pieragostini 80, Genova, 16151, Italy}
\author{Francesco Ferrari}
\affiliation{Quantum Computing Solutions, Leonardo S.p.A., Via R. Pieragostini 80, Genova, 16151, Italy}
\orcid{https://orcid.org/0000-0003-1543-2861}
\author{Daniele Dragoni}
\affiliation{Quantum Computing Solutions, Leonardo S.p.A., Via R. Pieragostini 80, Genova, 16151, Italy}
\affiliation{Hypercomputing Continuum Unit, Leonardo S.p.A., Via R. Pieragostini 80, Genova, 16151, Italy}
\author{Francesco Tudisco}
\affiliation{Gran Sasso Science Institute,  Viale Francesco Crispi 7, L'Aquila, Italy}
\affiliation{School of Mathematics and Maxwell Institute, University of Edinburgh, 
Peter Guthrie Tait Road,
EH9 3FD,
Edinburgh,
UK}
\orcid{https://orcid.org/0000-0002-8150-4475}

\begin{NoHyper}
\maketitle
\end{NoHyper}

\begin{abstract}
We study the problem of emergency operation center placement in disaster response, where a minimal number of hubs must be selected to ensure timely coverage of all affected locations. This task can be formulated as a minimum dominating set problem on a graph encoding reachability within a target response time. We propose a hybrid quantum–classical approximation framework that leverages neutral-atom quantum computers as independent set samplers. Candidate dominating sets are constructed from both small maximal independent sets and complements of large independent sets, and are subsequently refined via a lightweight classical procedure. We benchmark the approach on synthetic instances and realistic case studies, and implement it on the Fresnel quantum processor by Pasqal, solving instances of up to 100 nodes. Our results show that quantum-generated samples, despite hardware noise, enable near-optimal solutions of the placement problem. Overall, our results demonstrate that neutral-atom devices operating in analog mode can already be used to tackle graph optimization problems for real-world applications.
\end{abstract}

\section{Introduction}

Quantum computers~\cite{ezratty2025quantum} are being extensively explored as alternative computational platforms for specific industrially relevant tasks, with combinatorial optimization standing out as a primary application domain~\cite{giraldoquintero2026quantum,abbas2024challenges}. Several hardware architectures are currently under development, each natively supporting a certain operational paradigm~\cite{ladd2010qc,chae2024review}. Gate-based devices, realized across various physical platforms (e.g., superconducting circuits, trapped ions, spin qubits, and photonic systems), implement a digital, universal model of quantum computation~\cite{nielsen2010quantum}. Quantum annealers, by contrast, are analog machines specifically designed for combinatorial optimization, where the solution is encoded in the ground state of a problem Hamiltonian, reached by a controlled time evolution~\cite{Albash_2018,yarkoni2022annealer}.
Neutral atom platforms~\cite{Wintersperger2023,macri2026neutral} occupy an intermediate position in this rapidly evolving landscape: while hardware providers are progressively moving towards gate-based implementations~\cite{bluvstein2024logical}, their most mature use today is in analog mode~\cite{henriet2020neutralatoms, browaeys2020rydberg, bluvstein2021rydberg}. In this regime, the Rydberg blockade~\cite{lukin2001dipole, urban2009rydberg, saffman2010rydberg} mechanism naturally enforces independent-set constraints on the qubit register~\cite{Pichler2018MIS}, a class of constraints that arises in numerous graph-based combinatorial problems of practical interest. Because the constraint is imposed throughout the time evolution of the system, it can be exploited to design adiabatic algorithms that natively explore the space of independent sets~\cite{ebadi2022quantumMIS}.

Leveraging this architectural feature of neutral-atom platforms, we explore a hybrid quantum-classical algorithm for a graph domination problem arising in Emergency Management, namely the strategic placement of Emergency Operation Centers (EOCs) such as field hospitals and Search and Rescue (SAR) hubs in response to a natural disaster (e.g., landslides, floods, earthquakes). EOCs must be positioned so that every affected community can be reached quickly, while ensuring efficient allocation and deployment of limited available resources. This trade-off can be mathematically modeled as a minimum dominating set (mDS) problem, where the graph represents which candidate locations are reachable from one another within a predetermined maximum response time. To date, the application of quantum approaches to graph domination problems remains limited, with only a few studies~\cite{e26121057, Li2026} employing the Quantum Approximate Optimization Algorithm (QAOA)~\cite{farhi2014quantumapproximateoptimizationalgorithm}.

The hybrid quantum-classical algorithm proposed in this work exploits the connection between graph domination problems and independent sets (ISs)~\cite{marathe1994udg, haynes1998fundamentals, GODDARD2013839, CHO2023341} as an alternative computational route to solve the mDS on a graph. Independent sets are generated by sampling from a neutral-atom quantum processing unit (QPU) operated in analog mode~\cite{henriet2020neutralatoms} and inserted in a classical pipeline to obtain candidate minimum dominating sets, through a greedy removal of redundant nodes and local replacements. This yields an IS-to-mDS framework in which domination is not enforced directly on the QPU, but instead recovered through classical post-processing of ISs that can be natively sampled.

In this respect, the main contributions of this work are threefold. First, we formulate EOC placement as an mDS problem on travel-time graphs. Second, we introduce an IS-to-mDS approximation framework that leverages neutral-atom IS sampling to build good-quality dominating-set candidates, without imposing domination directly. Third, we benchmark the resulting hybrid workflow through experiments on the Fresnel QPU by Pasqal~\cite{henriet2020neutralatoms}, leveraging up to $100$ neutral atoms to tackle realistic EOC placement instances.

\begin{figure*}[t]
    \centering
    \begin{subfigure}[t]{0.49\linewidth}
        \centering
        \includegraphics[width=\linewidth]{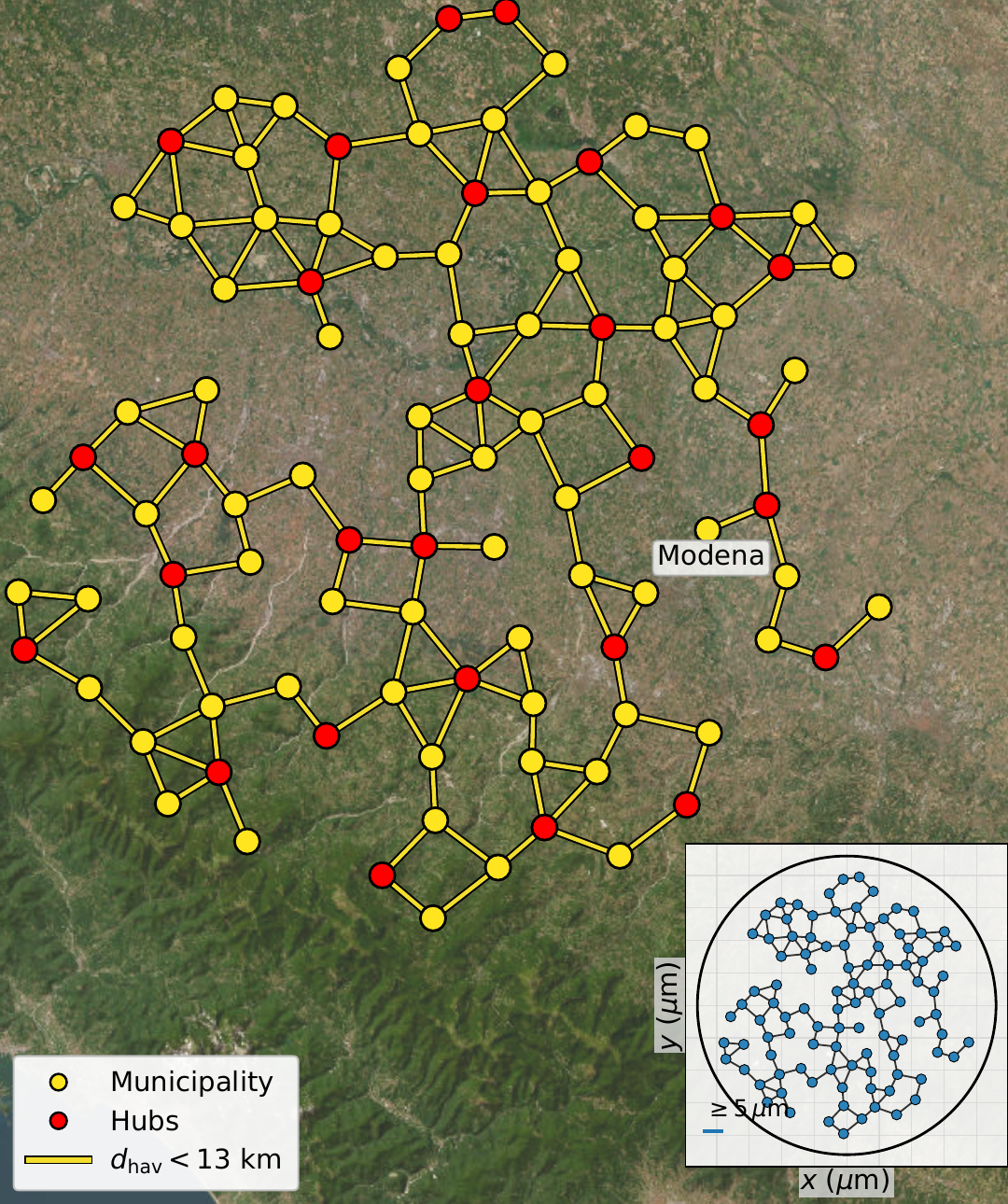}
    \end{subfigure}
    \hfill
    \begin{subfigure}[t]{0.49\linewidth}
        \centering
        \includegraphics[width=\linewidth]{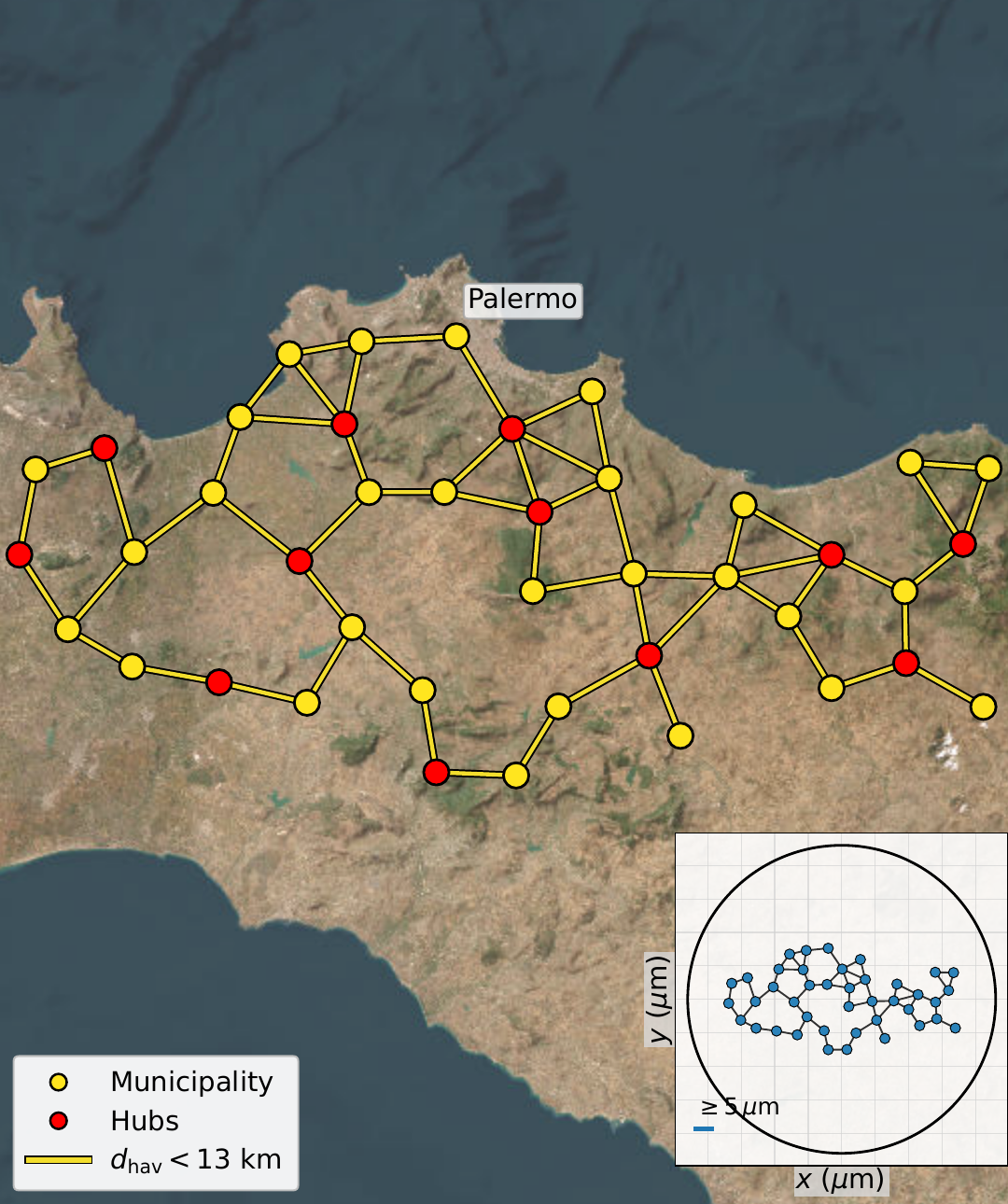}
    \end{subfigure}
    \caption{Case studies of emergency hub placement in the Po Valley area near Modena (left) and in north-western Sicily, including Palermo (right). Graph nodes represent municipalities that are candidate locations for the placement of EOCs. Edges connect pairs of locations which lie within the given travel time $T_{\rm max}$. Red nodes denote the optimal solution of the mDS problem~\eqref{eq:mds_obj} on these graphs. The bottom-left insets of both panels show the embedding of each problem graph in the neutral-atom QPU chamber, corresponding to the associated atomic register.}
    \label{fig:realistic_use_cases}
\end{figure*}

\section{EOC placement in disaster response scenarios}\label{sec:emergency-hub}

In the response phase following natural disasters such as floods or earthquakes, EOCs function as forward deployment bases for SAR assets, e.g. field medical units, ground rescue teams, unmanned aerial vehicles, helicopters. We tackle the problem of the strategic placement of these critical response facilities, which should be positioned so that first responders can reach the affected areas quickly and efficiently. 

In this work, we assume that reconnaissance is performed by drones and rescue operations are carried out by helicopters. Therefore, SAR hubs must be positioned so they can function both as drone launch sites for early-phase search operations and as bases for helicopter deployment. Small unmanned aerial vehicles (UAVs) can be deployed within minutes to survey affected areas, particularly where road access is uncertain or obstructed~\cite{7807176, GRIFFIN2014264}. The data collected by UAVs allow operators to identify potentially isolated individuals, verify incident reports, and assign response priorities based on observable conditions~\cite{MOHDDAUD202230}.

In practice, EOCs must be positioned so that emergency teams are able to reach every affected community within a strictly bounded response time $T_{\rm max}$, reflecting both operational requirements and the need for timely life-saving intervention. However, available resources, such as rescue teams, ground/aerial vehicles, and field equipment, are inherently limited and must not be excessively fragmented across a large number of locations, as this would reduce operational efficiency. This requires identifying the smallest subset from a large set of candidate locations at which SAR hubs can be positioned to ensure full coverage of the disaster-affected area.

\subsection{Minimum Dominating Set}

To model the EOC placement problem, we define a proximity graph $G=(V,E)$. Graph nodes represent populated areas (e.g., towns and cities) that form the pool of candidate facility locations for the SAR hubs. An edge connects pairs of nodes if the travel time between the two locations is below $T_{\rm max}$. The objective is to select a minimal subset of locations to host SAR hubs such that every location either hosts a hub or is adjacent to one. This corresponds to finding an mDS of $G$.

We introduce a binary decision variable $y_j\in\{0,1\}$ for each node $j \in V$, where $y_j=1$ if a hub is placed at node $j$. The mDS problem is
\begin{align}
\min\ & \sum_{j\in V} y_j \label{eq:mds_obj}\\
\text{s.t. } \quad &  y_i + \sum_{j\in {\cal N}(i)} y_j \ge 1 \qquad \forall i\in V \label{eq:domination}\\
& y_j\in\{0,1\}\qquad \forall j\in V, \label{eq:binary}
\end{align}
where ${\cal N}(i)$ denotes the set of neighbors of node $i$. \Cref{eq:domination} enforces the dominating set constraint, i.e., ensures that each municipality is either selected as a hub or covered by at least one neighboring hub within $T_{\rm max}$.

The mDS problem is NP-hard~\cite{garey1979computers}, via polynomial-time reduction from \textsc{Set Cover}~\cite{Karp1972,korte12}. As a result, exact traditional methods scale poorly~\cite{10.1145/1552285.1552286}, and practical approaches typically rely on approximations and heuristics, including local search and greedy degree- or centrality-based selection strategies~\cite{app14209251,ijcai2018p210,haraguchi2019efficientlocalsearchminimum,CASADO202341,ZHU2024111950,ijcai2025p997}. Exact linear-time methods are known only for restricted graph classes, such as the bloom graphs~\cite{Dharmakkan2022}. The mDS problem has also several applications in the fields of telecommunication and network analysis~\cite{GODQUIN2020102640,8233992}.

\subsection{Case studies and graph construction}

For our case study, we consider two hazard-exposed regions in Italy: a subregion of the Po Valley, spanning the provinces of Bologna, Modena, and Reggio Emilia, which has historically been affected by floods~\cite{cremonini2024floods_emilia_romagna}, and an area centered on Palermo, distributed across a coastal-to-inland corridor in north-western Sicily. We include $|V|=100$ municipalities within an approximately $40$–$45$ km neighborhood around Modena, and $|V|=41$ municipalities in Sicily, extending up to $\sim 105$ km from Palermo. 

We approximate travel times between locations using straight-line (airline) distances, assuming reconnaissance is performed by drones and rescue operations by helicopters. This reflects realistic SAR scenarios, where natural disasters often leave roads blocked, infrastructure damaged, and ground access unreliable or unsafe. To construct a representative and operationally plausible scenario, we adopt a simple set of assumptions regarding response operations. We assume a $10$-minute response-time target and an average transit speed of $150$ km/h over $5$ minutes of flight, after fixed overhead for activation and landing, yielding a one-way coverage radius of about $13$ km. We thus draw edges between municipalities that lie at a distance lower than this threshold. Coordinates of the municipalities are taken from \textit{openpolis/geojson-italy}~\cite{openpolis_geojson_italy}.

Both realistic scenarios considered in this work are depicted in \Cref{fig:realistic_use_cases}, in which candidate EOC positions are emphasized in yellow, while the red nodes indicate those selected by solving the mDS, guaranteeing fast simultaneous intervention in all nodes departing from the selected points.

\begin{table*}[t]
\centering
\begin{tabular}{@{}lll@{}}
\toprule
{Concept} & {Definition} & {Property} \\ \midrule
Independent Set (IS) & No adjacent nodes in the set & $y_i + y_j \leq 1\; \quad \forall \; i \in V \land j \in {\cal N}(i)$ \\
Dominating Set (DS) & Each node is in the set or covered & $y_i + \sum_{j\in {\cal N}(i)} y_j \ge 1 \quad \forall i\in V$  \\
Maximal Independent Set (mIS) & Independent and not extendable & Every mIS is a DS \\
$\overline{\mathrm{IS}}$ & Complement of an IS & Connected $G$ $\Rightarrow \overline{IS}$ is a DS \\
Minimum Dominating Set (mDS) & Smallest dominating set & --\\
\bottomrule
\end{tabular}
\caption{Summary of key graph concepts related to independent and dominating sets for a graph $G=(V,E), |V| \ge 2$.}
\label{tab:graph-concepts}
\end{table*}

\section{IS-based heuristics for mDS problems}\label{sec:method}

\begin{figure*}[t]
\centering
\begin{adjustbox}{width=2.1\columnwidth}
\begin{tikzpicture}[
  font=\Large,
  transform shape,
  node distance=5mm and 8mm,
  >={Stealth[length=3mm]},
  box/.style={
    rounded corners=2mm,
    draw=black!60,
    thick,
    align=center,
    minimum width=30mm,
    minimum height=8mm,
    inner sep=4pt,
    fill=black!2,
    blur shadow
  },
  io/.style={
    trapezium,
    trapezium left angle=70,
    trapezium right angle=110,
    draw=black!60,
    thick,
    align=center,
    minimum width=30mm,
    minimum height=8mm,
    inner sep=4pt,
    fill=black!2,
    blur shadow
  },
  arrow/.style={->, thick, draw=black!70},
  lab/.style={midway, fill=white, inner sep=2pt, text=black!80}
]




\node[io] (graph) {Input graph \\ $G=(V,E)$};

\node[box, below right=of graph] (pool) {Sample mISs\\
$\mathcal{I}=\{I_1,\dots,I_M\}$};

\draw[arrow] (graph) |- (pool.west);

\node[box, above right=4mm and -9mm of pool, yshift=6mm] (smallmis)
{Keep smallest mISs\\
$\ell_{\min}=\min_{I\in\mathcal{I}} |I|$\\
$\mathcal{D}_{\mathrm{mIS}}=\{I: \ I\in\mathcal{I}, \ |I|=\ell_{\min}\}$};


\node[box, below right=1mm and -12.5mm of pool, yshift=-10mm] (largemis)
{Take complements of largest mISs\\
$\ell_{\max}=\max_{I\in\mathcal{I}} |I|$\\
$\mathcal{D}_{\overline{\mathrm{IS}}}=\{V\setminus I: \ I\in\mathcal{I}, \ |I|=\ell_{\max}\}$};

\node[box, right=40mm of pool] (ref)
{Greedy refinement: \\ 
(i) prune redundant nodes\\
(ii) 2-neighborhood sweep};

\node[io, right=of ref] (select) {Select best over \\
refined candidates\\
$D^\star=\underset{D \in \mathcal{R}}{\arg\min} |D|$};


\draw[arrow] (pool) |- (smallmis.west);
\draw[arrow] (pool) |- (largemis.west);

\draw[arrow] (largemis.east) -| (ref.south);

\draw[arrow] (smallmis.east) -| (ref.north);

\draw[arrow] (ref) -- (select);

\end{tikzpicture}
\end{adjustbox}
\caption{Two-branch workflow for approximating the minimum dominating set (mDS) from sampled maximal independent sets (mIS), obtained either by classical or quantum methods.}
\label{fig:mds-workflow}
\end{figure*}
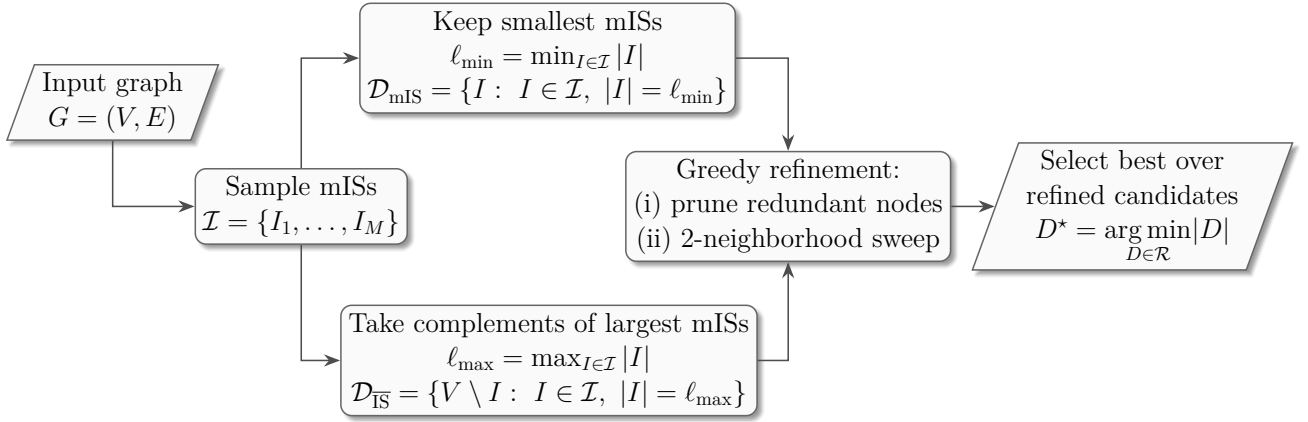

In this section, we describe the proposed workflow to construct high-quality and scalable approximations for the mDS problem. The graph-theoretic notions and properties exploited by the algorithm are summarized in \Cref{tab:graph-concepts}. We begin with some preliminary observations that highlight the connection between dominating sets (DSs) of a graph and independent sets (ISs), i.e., sets of a graph that do not contain pairwise adjacent nodes. First, we observe that every maximal independent set (mIS)\footnote{Not to be confused with the \textit{maximum} independent set, which is the largest IS of a graph, and hence the largest among all mIS.}, i.e., an IS that cannot be further enlarged by adding any additional node, is also a DS in an arbitrary graph. This property follows from maximality: if a set is independent and maximal, then every node outside of it has a neighbor in it, otherwise, it could be added to the set without breaking independence. This allows us to find mDS candidates among \textit{small} mISs. Second, we note that, in connected graphs $G=(V,E)$ with $|V| \ge 2$, the complement of any IS, here denoted as $\overline{\mathrm{IS}}$, is itself a DS.
Indeed, given any vertex $v \in V$, and an independent set $S$ of $G$, if $v\in V\setminus S = \overline{\mathrm{S}}$, then it belongs to the proposed dominating set; instead, if $v \in S$, then $v$ has at least one neighbor because $|V| \ge 2$ and $G$ is connected. Since $S$ is independent, none of the neighbors of $v$ can belong to $S$. Therefore, in this case, at least one neighbor of $v$ belongs to $\overline{\mathrm{S}}$, so $v$ is dominated by $\overline{\mathrm{S}}$. This directly implies that $\overline{\mathrm{S}}$ is a dominating set.

The procedure we propose is based on using mISs as valid candidates for mDS, which are subsequently refined towards minimality. Specifically, it begins by randomly generating a pool of $M$ mISs, $\mathcal{I} = \{I_1,\dots,I_M\}$, either by a classical sampler, e.g. the one provided in the NetworkX Python library~\cite{SciPyProceedings11}, or using the neutral-atom quantum hardware, as discussed in \Cref{sec:qpu}. After a pool of mIS is obtained, as illustrated in \Cref{fig:mds-workflow}, the workflow splits in two branches, both of which generate candidate DSs to be refined by a greedy procedure.

The first branch is motivated by the observation that any mIS is a DS. Therefore, small mIS instances are natural proxies for small DSs, so from the sampled pool $\mathcal{I}$ we retain the smallest mISs:
\begin{equation}
\mathcal{D}_{\mathrm{mIS}} = \{\, I \in \mathcal{I}\;:\; |I|=\ell_{\min}\,\},
\end{equation}
where $\ell_{\min} = \min_{I\in \mathcal{I}} |I|$. Each $D \in \mathcal{D}_{\mathrm{mIS}}$ is a feasible dominating set.

The second branch builds approximate mDS candidates from complements of large mISs. Specifically, we identify the complements of the largest mISs in $\mathcal{I}$, i.e.
\begin{equation}
\mathcal{D}_{\overline{\mathrm{IS}}}=\{V\setminus I: \ I\in\mathcal{I}, \ |I|=\ell_{\max}\}
\end{equation}
where $\ell_{\max} = \max_{I\in \mathcal{I}} |I|$. Since, as noted above, in connected graphs $\overline{I}$ is a DS whenever $I$ is a mIS, then $\mathcal{D}_{\overline{\mathrm{IS}}}$ yields a second family of feasible DS candidates.

Both $\mathcal{D}_{\mathrm{mIS}}$ and $\mathcal{D}_{\overline{\mathrm{IS}}}$ contain valid dominating sets, but they are not necessarily minimal. In particular, the $\mathcal{D}_{\overline{\mathrm{IS}}}$ calculated this way is extremely redundant. The total candidate sets are given by $\mathcal{D}=\mathcal{D}_{\mathrm{mIS}} \ \cup \ \mathcal{D}_{\overline{\mathrm{IS}}}$. We then apply a lightweight two-stage local refinement procedure to each element in $\mathcal{D}_{\mathrm{mIS}}$ and $\mathcal{D}_{\overline{\mathrm{IS}}}$, yielding the refined set of dominating sets $\mathcal{R}_{\mathrm{mIS}}$ and $\mathcal{R}_{\overline{\mathrm{IS}}}$, respectively. The first step consists in a greedy redundancy pruning: nodes in the candidate set $D \in \mathcal{D}$ are examined in increasing-degree order and removed whenever domination is preserved. After that, we apply a $2$-neighborhood sweep inspired by efficient local-search rules for mDS~\cite{haraguchi2019efficientlocalsearchminimum}: running over all nodes $i\notin D$, if $i$ is adjacent to exactly two nodes $j,k\in D$, we propose the new set $D'=(D\setminus\{j,k\})\cup\{i\}$ and accept it if $D'$ is still a DS. The smallest DSs from the two branches are
\begin{align}
 D_{\mathrm{mIS}}^\star &= \underset{{D \in \mathcal{R}_{\mathrm{mIS}}}}{\arg\min}  |D|, \\
 D_{\overline{\mathrm{IS}}}^\star  &=  \underset{{D \in \mathcal{R}_{\overline{\mathrm{IS}}}}}{\arg\min}  |D|. 
\end{align}
The overall best approximation for the mDS is the smallest of the two sets above, denoted as $D^\star$.

\section{Rydberg atoms mIS sampler}\label{sec:qpu}

The procedure outlined so far can use both quantum and classical sampling methods to generate the pool of candidate ISs. In this work, we focus primarily on an approach that uses samples obtained from a neutral-atom quantum computer.

neutral-atom QPUs utilize laser-trapped atomic arrays in programmable geometries, controlling their quantum evolution via laser pulses. In these systems, the Rydberg blockade interaction~\cite{lukin2001dipole, urban2009rydberg, saffman2010rydberg} prevents simultaneous excitation of nearby atoms within a characteristic blockade radius. Atoms can thus be spatially arranged to represent the nodes of a desired graph, with edges connecting pairs of atoms separated by less than the blockade radius, for which simultaneous excitation is energetically suppressed. In the regime where the blockade constraint is effectively enforced (i.e., interaction energies dominate over driving), the accessible low-energy subspace is restricted to configurations corresponding to ISs of the graph. This property has been exploited, for instance, to tackle the maximum independent set problem~\cite{pichler2018optimization, ebadi2022quantumMIS,cazals2025identifyinghardnativeinstances,7dkh-crjj, Dalyac2024, wurtz2024industry, rava2025benchmarkingneutralatombasedquantum}. In this work, we leverage the Rydberg blockade to define a mIS sampler for the mDS workflow described in \Cref{sec:method}.

To represent the graph $G$ of the mDS problem, we embed its structure by mapping its nodes to the spatial coordinates of the atoms. For our specific use case concerning the strategic placement of EOCs for SAR operations using drones and helicopters, $G$ is natively a unit-disk graph~\cite{marathe1994udg}, since the edges connect pairs of nodes located below a certain straight-line distance. The atomic embedding is thus achieved simply by rescaling the geographic coordinates, without the need for complex embedding strategies required for more general graphs~\cite{BREU19983, vercellino2022neural, de_Correc_2025}. The time-evolution of the atoms is governed by the Rydberg Hamiltonian 
\begin{equation} H(t)=\Omega(t)\sum_{i \in V}\sigma_i^x-\delta(t)\sum_{i \in V} n_i+\sum_{i<j \in V} \frac{C_6}{R_{ij}^6}\,n_i n_j. \label{eq:rydbergH}
\end{equation}
The Hamiltonian is divided into user-controlled and intrinsic interaction terms. The externally controlled parameters are the time-dependent Rabi frequency $\Omega(t)$ and detuning $\delta(t)$, modulating the atomic Pauli-$X$ ($\sigma_i^x$) and number ($n_i$) operators. In contrast, the density–density repulsive term depends only on the interatomic distance $R_{ij}$ and accounts for the van der Waals interaction between atoms. Its strength is set by the van der Waals coefficient $C_6$. This interaction is responsible for the Rydberg blockade effect.

We employ an analog time-evolution protocol to generate ISs of the graph. For this purpose, the atoms are all initialized in their ground state $|0\rangle$ and evolved with suitable choices of Rabi frequency and detuning. At the end of the protocol, the atoms are measured, yielding a binary bitstring.
The nodes corresponding to atoms in state $|1\rangle$ in a given bitstring form the set $S$ of candidate IS nodes. Indeed, provided the laser pulses $\Omega(t)$ and $\delta(t)$ are properly designed, the strong interaction $V_{ij}$ suppresses simultaneous excitations of nearby atoms, biasing the system towards final bitstrings that represent valid ISs of $G$. In practice, however, raw quantum samples can still violate the IS condition and contain residual conflicts (adjacent excited nodes caused by hardware noise or non-ideal blockade along the adiabatic protocol). Furthermore, even if a suitable choice of the adiabatic schedule can push the system towards large ISs, the maximality condition is not guaranteed.

Therefore, to generate a suitable candidate set $\mathcal{I}$, we apply a deterministic correction routine to correct the samples and transform them into mISs, which can then be used to feed the workflow of \Cref{fig:mds-workflow}. We first remove conflicting selections to enforce the IS condition, and then greedily extend the resulting IS until maximality is reached. The samples are corrected in two opposite ways, motivated by the splitting of the algorithm into the branches producing $\mathcal{D}_{\mathrm{mIS}}$ and $\mathcal{D}_{\overline{\mathrm{IS}}}$. While the branch producing $\mathcal{D}_{\mathrm{mIS}}$ aims for smaller mISs, the complement-based branch yielding $\mathcal{D}_{\overline{\mathrm{IS}}}$ aims for larger mISs, which inherently imply smaller complements. The greedy extension utilizes two degree-based tie-breaking strategies: to favor larger mISs, we prioritize adding nodes with fewer neighbors, whereas to target smaller mISs, we prioritize nodes with higher degrees. 
This procedure ensures that the resulting samples respect independence constraints and the maximality condition, hence are a suitable set $\mathcal{I}$.

\section{Results}

We evaluate the performance of the proposed algorithm for the mDS problem applied to optimal EOC placement~\eqref{eq:mds_obj}. We first present an extensive benchmark on synthetic instances, in which we emulate the quantum algorithm on classical hardware using the matrix-product state backend of the Pulser library by Pasqal~\cite{Silv_rio_2022}. Calculations are carried out on \textit{davinci-1}, the proprietary supercomputer of Leonardo S.p.A. We then discuss the results obtained on the Fresnel neutral-atom QPU by Pasqal for the EOC placement case studies in the Modena and Palermo regions, see \Cref{fig:realistic_use_cases}.

\begin{figure}[t]
    \centering
    \includegraphics[width=\columnwidth]{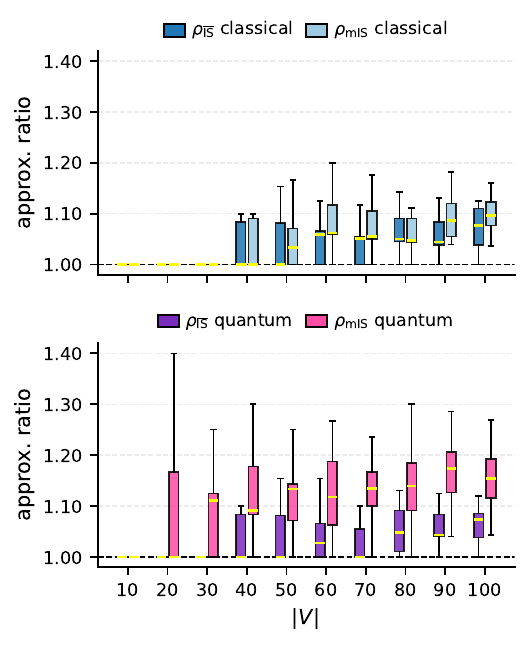}
    \caption{Boxplot of the approximation ratios found by each branch of the proposed algorithm, $\rho_{\rm mIS}$ and $\rho_{\overline{\rm IS}}$ [see \Cref{eq:approx-ratio}], as a function of graph size $|V|$. We collect statistics on $30$ test instances of increasing size up to $|V|=100$ on randomly generated unit-disk graphs. The upper panel shows results obtained by sampling mISs classically, while the lower panel shows those obtained with the emulated quantum sampler.}
    \label{fig:UDG-approx-ratio}
\end{figure}

\begin{figure}[t]
    \centering
    \includegraphics[width=\columnwidth]{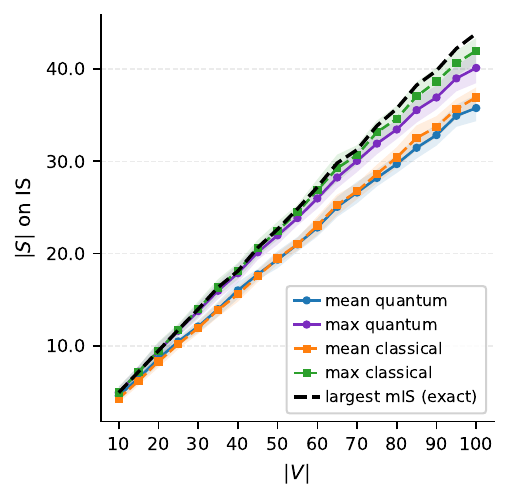}
    \caption{Mean and maximum sizes of the independent sets obtained with the quantum and classical samplers. We show their average (dots) and the standard deviation (shaded area) across the instances used for the statistical experiments summarized in \Cref{fig:UDG-approx-ratio}. The dashed black line indicates the size of the largest mIS, averaged over the various instances.}
    \label{fig:mean-and-max-IS-size}
\end{figure}

\begin{figure*}[t]
    \centering
    \includegraphics[width=\linewidth]{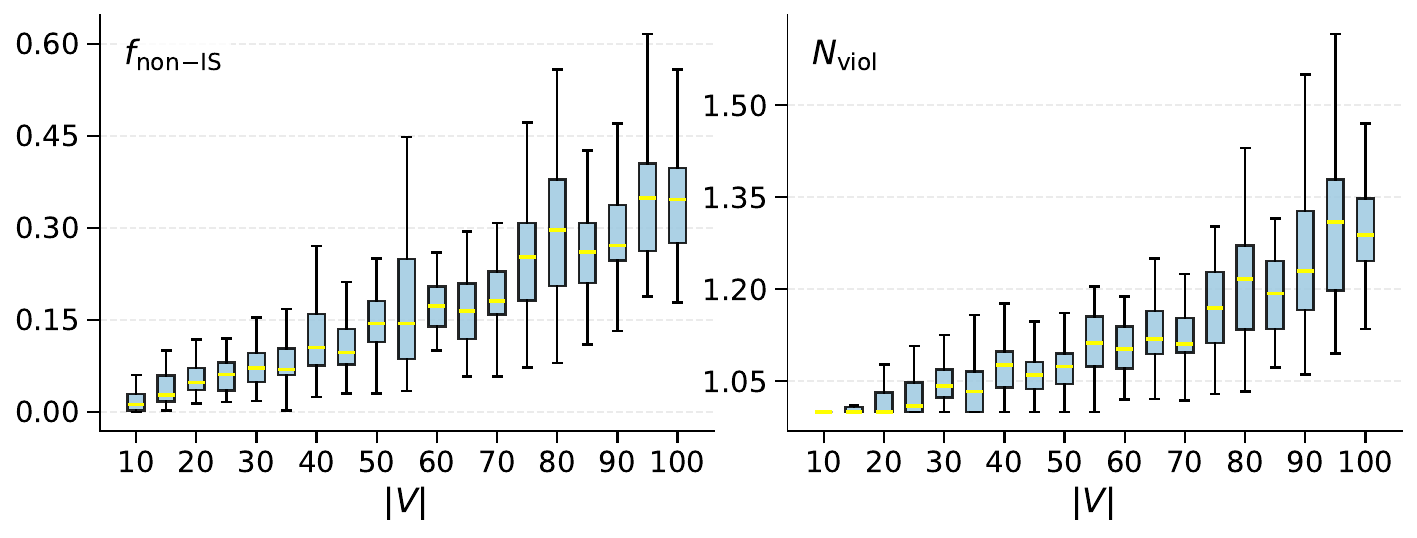}
    \caption{Left panel: frequency of invalid (i.e., non-IS) samples $f_{\neg {\rm IS}}$ among raw bitstrings from the emulated quantum algorithm. Right panel: severity of the IS violation among non-IS samples, as measured by the number of violating edges $N_{\rm viol}$. Statistics are collected over the same instances used for \Cref{fig:UDG-approx-ratio}.}
    \label{fig:freq-notIS-and-severity}
\end{figure*}

\subsection{Benchmarks on synthetic instances}\label{subsec:synthetic-tests}

We generate synthetic mDS instances as random unit-disk graphs and assess the performance of the proposed algorithm. We compare a fully classical pipeline in which the set ${\cal I}$ is generated by the NetworkX mIS sampler and a quantum scheme emulating the use of a neutral-atom computer, as explained in \Cref{sec:qpu}. As an exact reference, exact mDSs are computed with the branch-and-bound method implemented in the IBM CPLEX library~\cite{cplex127manual}. For each of the two branches of the algorithm, we evaluate the approximation ratio 
\begin{align}\label{eq:approx-ratio}
\rho_{\mathrm{mIS}/\overline{\mathrm{IS}}} = \frac{\left|D^\star_{\mathrm{mIS}/\overline{\mathrm{IS}}}\right|}{ |\mathrm{mDS}|},
\end{align}
where the denominator is the exact mDS. 

For the quantum computing algorithm, the pulse schedule consists of three stages over a total evolution time $t_{\rm tot}$: an amplitude ramp $0\!\to\!\Omega_{\max}$ at fixed detuning $\delta_0$ over $t_{\mathrm{rise}}=t_{\rm tot}/6$, a detuning sweep ${\delta_0\!\to\!\delta_f}$ at fixed amplitude $\Omega_{\max}$ over ${t_{\mathrm{sweep}}=t_{\rm tot}/2}$, and a final amplitude ramp ${\Omega_{\max}\!\to\!0}$ at fixed detuning $\delta_f$ over $t_{\mathrm{fall}}=t_{\rm tot}/3$. We take $\Omega_{\max}=2\pi$ rad/$\mu$s and rescale the positions of the nodes to atomic units such that edges are compatible with a Rydberg blockade radius $R_{b}=\left(\frac{C_6}{\hbar \Omega_{\max}}\right)^{1/6}$. We fix the initial detuning to $\delta_0=-3\Omega_{\max}$~\cite{Silv_rio_2022}, and optimize $t_{\rm tot}$ and $\delta_f$ through a grid search on a representative instance with $|V|=50$ nodes. The aim is to identify a schedule that maximizes the fraction of valid IS samples, while favoring large sets which are more likely to be mISs. The resulting parameters are then reused across the full benchmark, since all instances share similar density and local interaction structure by construction.

Boxplots for the approximation ratio at variable problem size are reported in \Cref{fig:UDG-approx-ratio}. We collect statistics over $30$ instances per size, using $M=500$ mIS samples per instance (corresponding to $500$ measurement shots in the quantum approach). Both the classical and quantum approaches reach approximation ratios close to $1$ across the full benchmark, including the largest $|V|=100$ instances, indicating near-optimal performance. The $\overline{\mathrm{IS}}$ branch appears slightly more stable and more concentrated around $\rho\approx 1$, while the mIS branch shows larger dispersion and worse performance. Although variability increases moderately with graph size for all methods, the degradation remains limited: in the worst case, the approximated $D^\star$ deviates at most of $3$ nodes from the optimal mDS solution across the whole range of $|V|$ considered here. Notably, the quantum variants of the algorithm remain close to their classical counterparts, showing that neutral-atom sampling, combined with correction and refinement, provides good-quality mDSs. In particular, the $\overline{\mathrm{IS}}$ branch of the quantum algorithm slightly outperforms its classical counterpart. 

The quantum sampler based on neutral atoms is expected to yield bitstrings that satisfy the IS constraint with high probability. For the current application, we target raw bitstrings that are as close as possible to mISs, which, in the unit-disk graphs for EOC placement problems, are empirically observed to correspond to large ISs. While this can be achieved by tuning the protocol parameters, there is an intrinsic trade-off between maximizing IS sampling probability and obtaining large independent sets. Indeed, selecting a larger value for the parameter $\delta_f$ targets a larger IS size, but simultaneously increases the frequency of Rydberg blockade violations. 

To show that the selected parameters allow for sampling ISs of relatively large size, in \Cref{fig:mean-and-max-IS-size} we plot the average size of sampled ISs as a function of the graph size $|V|$. The plot compares the mean and maximum sizes of the ISs sampled by the classical mIS sampler and by the quantum sampler (before correction). In the classical case, all generated sets are ISs by construction. In the quantum case, statistics are collected only over raw bitstrings that are valid IS samples, excluding those that violate the IS condition and require classical correction. As expected, both quantities increase with $|V|$ for the two methods, demonstrating their capability to generate larger ISs as $|V|$ increases. Notably, for the selected analog evolution protocol, the neutral-atom-based sampler yields an IS-size distribution that closely matches that of the classical sampler. Both the quantum and classical samplers are capable of yielding high-cardinality ISs with a size close to that of the exact largest IS, as shown in \Cref{fig:mean-and-max-IS-size}. This suggests that, for the chosen $\delta_f$ and $t_{\rm tot}$ parameters, the quantum protocol effectively reproduces the target distribution of ISs. However, these considerations are meaningful only if the chosen protocol actually produces \textit{valid} ISs with high probability.

For this reason, we further characterize the distribution of the raw bitstrings generated by the quantum sampler (before corrections), by evaluating two additional quantities: the fraction of measured bitstrings that violate the IS constraint ($f_{\neg {\rm IS}}$), and, within those invalid bitstrings, the average number of edges causing the violation ($N_{\rm viol}$). The results are summarized by the boxplots in \Cref{fig:freq-notIS-and-severity}, which emphasizes how the neutral-atom QPU can reliably sample ISs: violations occur at a relatively low rate, and, when they do happen, they typically involve only a small number of conflicting edges. This shows that the quantum sampler is not merely acting as a generic random bitstring generator. Even in the presence of imperfect blockade and finite-time evolution protocol leading to possible IS violations, the method biases the output distribution toward large ISs, i.e. towards the regime needed by the proposed mDS algorithm.

\subsection{Results on quantum hardware for realistic instances}\label{subsec:realistic-tests}

We consider the realistic emergency instances of the EOC placement problem introduced in \Cref{sec:emergency-hub}, for the geographic areas around Modena and Palermo in Italy. For these case studies, we test our algorithm on the Fresnel neutral-atom quantum computer by Pasqal.

To sample ISs, we use the same adiabatic protocol described in the previous section. Since the geographic positions of municipalities already define valid unit-disk graphs, we simply rescale the coordinates to atomic units so that all nodes fit within the available space of Fresnel. The resulting unit-disk radius defines the blockade radius $R_b$, thus fixing the maximum frequency $\Omega_{\rm max}$ for our adiabatic protocol. The other parameters of the pulse schedule, $t_{\rm tot}$ and $\delta_f$, are optimized to identify operating regimes that simultaneously favor a high fraction of valid IS samples and large sampled sets. The parameter tuning is performed by a grid-based search for each of the two case studies, using numerical emulation on classical hardware.

\begin{figure}[t]
\centering

\begin{subfigure}[t]{0.95\linewidth}
    \centering
    \caption{Emulator}
    \includegraphics[width=\linewidth]{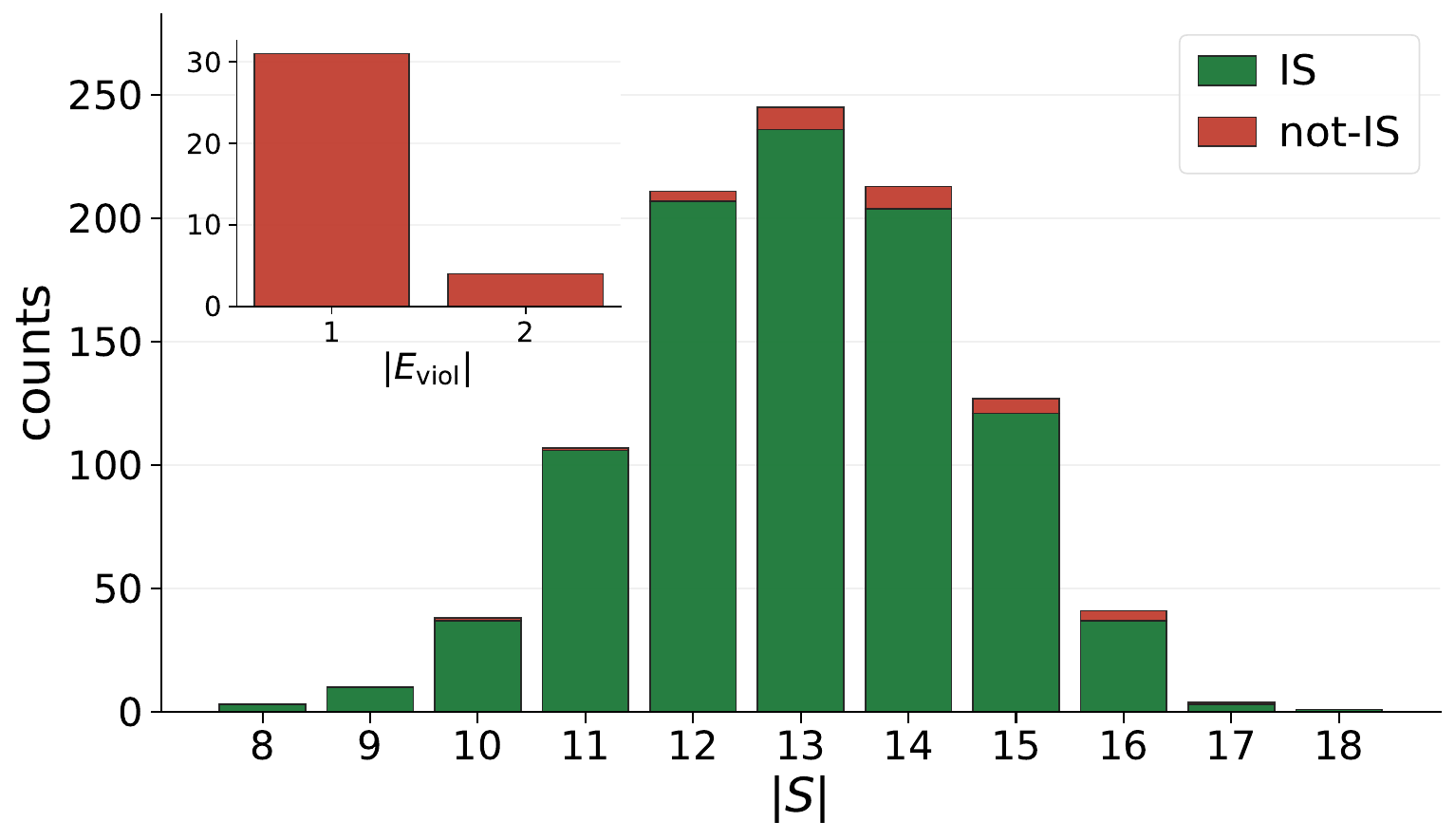}
    \label{fig:sicily-distribution-emulator}
\end{subfigure}

\vspace{0.5em}

\begin{subfigure}[t]{0.95\linewidth}
    \centering
    \caption{QPU}
    \includegraphics[width=\linewidth]{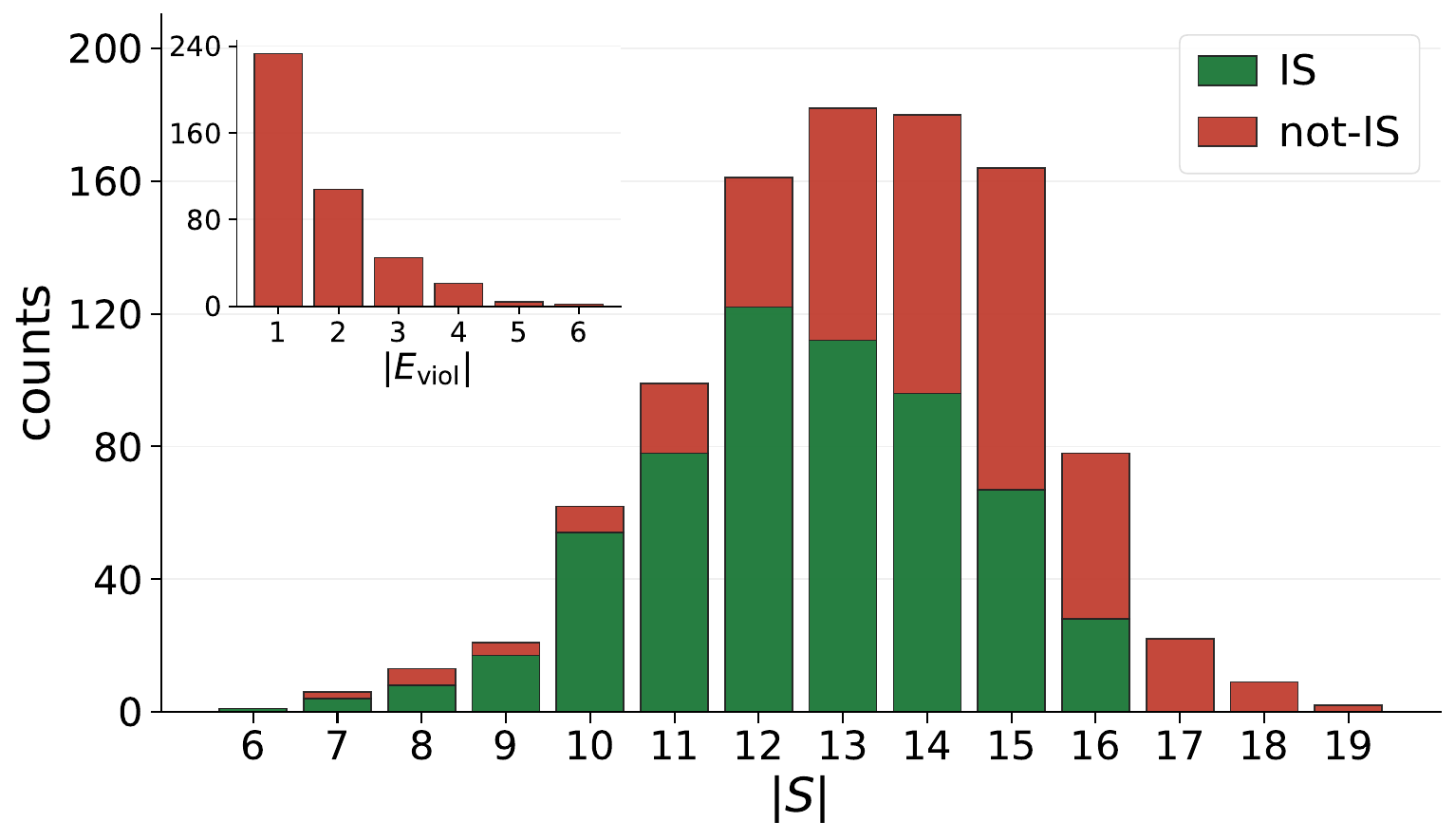}
    \label{fig:sicily-distribution-qpu}
\end{subfigure}

\caption{
Distribution of $1000$ raw quantum samples for the Palermo instance, plotted as a function of the size of sampled sets $|S|$. Valid ISs are shown in green, while non-IS samples are shown in red. The inset reports the severity of the violations as the distribution of the number of conflicting edges, i.e., edges connecting two selected nodes. Panel (a) and (b) refer to results in emulation and on Fresnel QPU, respectively.
}
\label{fig:sicily-distribution-1000-samples}
\end{figure}

Let us first discuss the results for the smaller instance in the Palermo area, consisting of $|V|=41$ candidate locations. The quantum algorithm is tested both in emulation and on the QPU, collecting ${M=1000}$ samples in each case. In emulation, $96.5\%$ of the samples turn out to be valid independent sets, while on the QPU this fraction decreases to $58.7\%$. This comparison is shown in \Cref{fig:sicily-distribution-1000-samples}, which reports the raw sample distributions by cardinality of the samples $|S|$ for both settings. Valid IS samples are shown in green, while samples violating the IS condition are shown in red. The inset further characterizes the infeasible samples by reporting the distribution of the number of violating edges $N_{\mathrm{viol}}$ among non-IS samples.

The emulated sampler produces ISs with sizes ranging from $8$ to $18$, whereas QPU valid IS samples range from $6$ to $16$. QPU samples of larger size violate the IS constraint, highlighting the difficulty of maintaining a perfect blockade on the QPU when targeting large ISs. In emulation, on the other hand, the distribution reaches configurations only one node smaller than the largest possible mIS on the graph, which has size $19$.

Although the fraction of samples that violate the IS constraint increases on the real device, the observed violations are still relatively mild and, once corrected, samples can serve as meaningful starting points for the subsequent mDS construction procedures. Among samples that do not satisfy the IS constraint, the average number of violating edges is $N_{\rm viol}\approx 1.1$ in emulation and $N_{\rm viol}=1.69$ on the QPU. Overall, the comparison highlights the expected performance gap between ideal emulation and current hardware execution. The structure of the sampled distribution is preserved, and violations remain limited. This suggests that the neutral-atom device is already sampling from relevant regions of the solution space, even in the presence of hardware imperfections.

For what concerns the final result of the mDS quantum pipeline, while the mIS branch yields a slightly suboptimal DS of $13$ nodes, the $\overline{\mathrm{IS}}$ branch retrieves the exact mDS solution with $12$ nodes shown in \Cref{fig:realistic_use_cases}. This can be compared to a run of the classical pipeline with $1000$ sampled mIS which also yields the optimal mDS.

\begin{figure}[t]
\centering

\begin{subfigure}[t]{0.95\linewidth}
    \centering
    \caption{Emulator}
    \includegraphics[width=\linewidth]{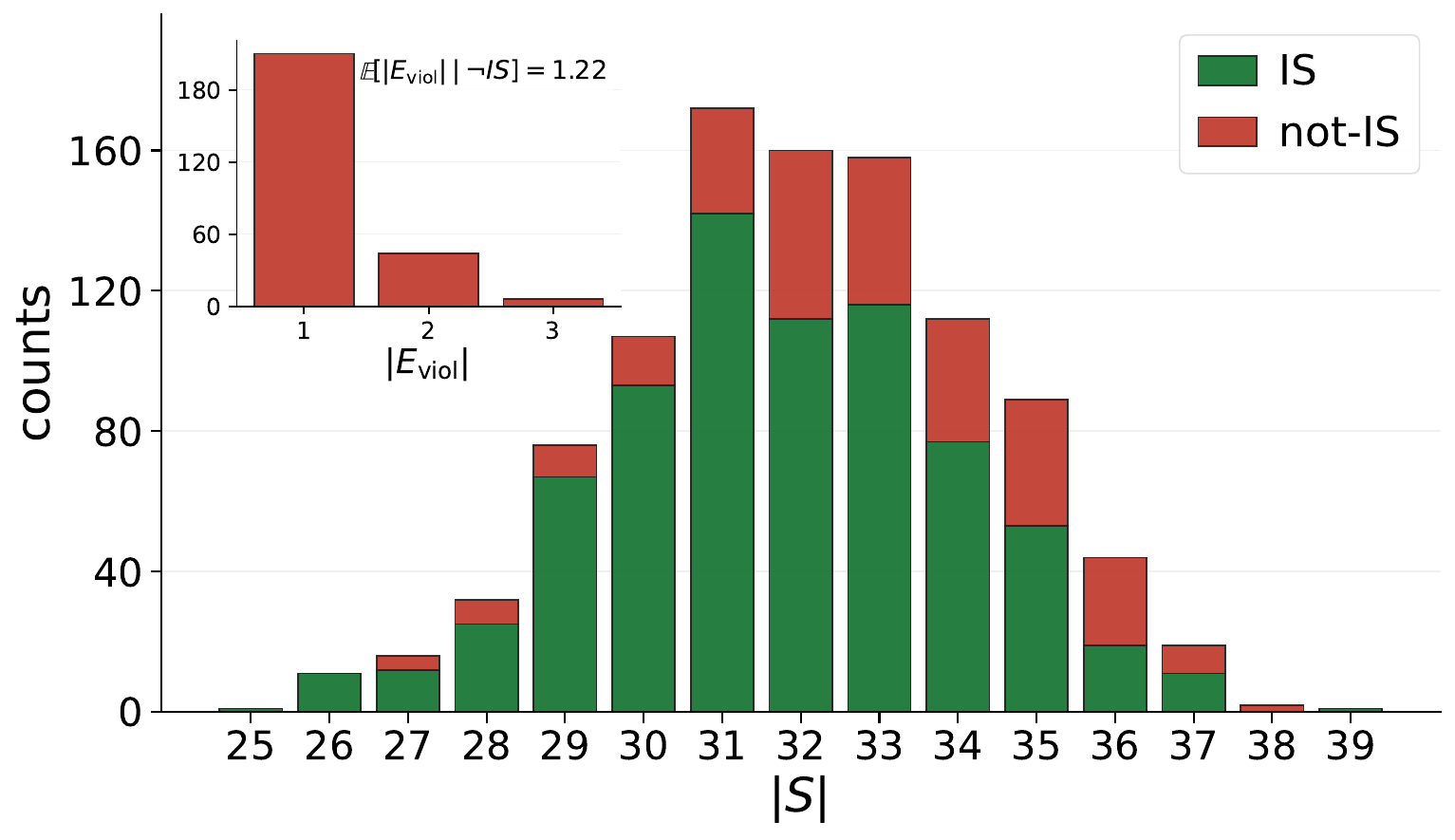}
    \label{fig:modena-distribution-emulator}
\end{subfigure}

\vspace{0.5em}

\begin{subfigure}[t]{0.95\linewidth}
    \centering
    \caption{QPU}
    \includegraphics[width=\linewidth]{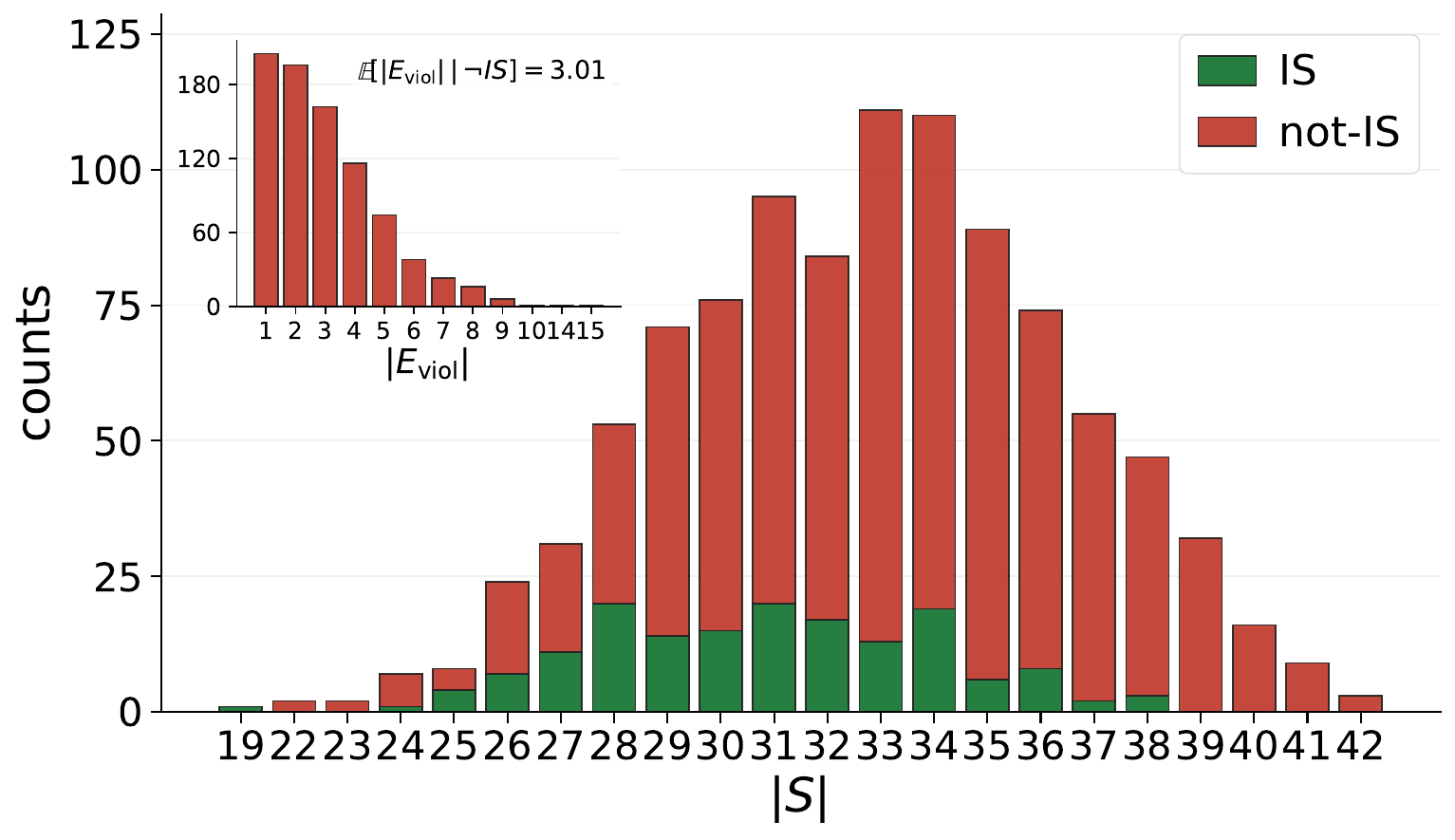}
    \label{fig:modena-distribution-qpu}
\end{subfigure}

\caption{
The same as \Cref{fig:sicily-distribution-1000-samples} but for the Modena instance.
}
\label{fig:modena-distribution-1000-samples}
\end{figure}

We tested the same quantum workflow on the EOC placement instance in the surroundings of Modena, represented by a graph with $|V|=100$ nodes. This case is particularly relevant because $100$ atoms correspond to the maximum register size available on Fresnel QPU. Compared to the previously discussed case, the hardware execution exhibits a stronger degradation with respect to emulation, as shown in \Cref{fig:modena-distribution-1000-samples}. In the emulated setting, the distribution reaches a $74 \%$ fraction of valid ISs, which drops to $16.1\%$ on the QPU. The severity of the violations increases on QPU, with invalid IS samples exhibiting an average of $N_{\rm viol} \approx 3.0$ violating edges, with peaks of $N_{\rm viol} = 15$ violations.

The degradation in QPU performance on the larger graph is consistent with the increased problem size. To embed the unit-disk graph on the device, we uniformly rescale the node coordinates by the minimum factor required for all atoms to fit within the Fresnel trapping region (a circle of radius $46$ $\mu$m). As the number of nodes grows, the required compression reduces the effective blockade radius $R_b$, thus increasing the maximum Rabi frequency $\Omega_{\rm max}$ compatible with it. Indeed, while for the $41$-nodes Palermo graph we obtain $\Omega_{\rm max}=3.9$ rad/$\mu$s, for the $100$-nodes Modena graph we get $\Omega_{\rm max}=4.5$ rad/$\mu$s. The latter frequency requires a longer optimal simulation time $t_{\rm tot}=5400$ ns, which is close to the maximum duration supported by Fresnel, making the protocol more susceptible to decoherence effects on QPU. Furthermore, we emphasize that our goal is to assess the capabilities of the current neutral-atom QPU on EOC placement instances that reflect realistic scenarios. Accordingly, the considered graphs are not artificial constructions designed to maximize the blockade effect, but are derived from geographical data and therefore exhibit nonuniform interatomic distances, which makes the enforcement of the blockade more challenging.

Although the current $100$-node instance hits the maximum register size of Fresnel, the QPU samples remain concentrated in a meaningful size range, and the infeasible samples are sufficiently close to valid configurations so that post-processing corrections can still enable the generation of high-quality mIS samples to be fed into the two branches of our algorithm.

\begin{table}[t]
\centering
\begin{tabular}{|c|c|cc|}
\hline
\multirow{ 2}{*}{} & & Palermo & Modena \\
\hline
\hline
Exact & mDS & 12 & 28 \\
\hline
\multirow{ 2}{*}{Classical} & $D^\star_{\mathrm{mIS}}$ & 12 & 30 \\
 & $D^\star_{\overline{\mathrm{IS}}}$  & 12 & 30 \\
\hline
\multirow{ 2}{*}{Emulation}  & $D^\star_{\mathrm{mIS}}$ & 13 & 31 \\
 & $D^\star_{\overline{\mathrm{IS}}}$ & 12 & 29 \\
\hline
\multirow{ 2}{*}{Fresnel QPU} & $D^\star_{\mathrm{mIS}}$  & 13 & 31 \\
& $D^\star_{\overline{\mathrm{IS}}}$ & 12 & 29 \\
\hline
\end{tabular}
\caption{
Summary of the results for the EOC placement case studies of Palermo ($|V|=41$ nodes/atoms) and Modena ($|V|=100$ nodes/atoms) on the Italian territory. We show the exact mDS obtained with the branch-and-bound algorithm, and results obtained with our pipeline using samples extracted from the classical sampler, the emulated QPU and the real Fresnel QPU. }
\label{tab:final_mds}
\end{table}

The results obtained on the Fresnel QPU for the two case studies are reported in \Cref{tab:final_mds}, together with the corresponding results in emulation using Pulser and from the classical pipeline. Specifically, on the real QPU, the $\overline{\mathrm{IS}}$ and mIS branches produce mDSs of sizes $29$ and $30$, respectively, compared to the exact mDS of size $28$. It is important to note that although for this $100$-node instance the fraction of valid IS samples drops on the QPU, the final DS obtained after correction and refinement requires only one additional hub. This suggests that the proposed mDS workflow can remain effective even when the quality of the raw samples deteriorates, indicating a degree of robustness to moderate noise and IS violations, whose impact is expected to diminish with future hardware improvements. At the same time, the results suggest that, with the considered QPU, realizing an effective blockade becomes increasingly challenging as the instance size grows, particularly for realistic, non-uniform geometric graphs.

\section{Conclusions}\label{sec:conclusions}

In this work, we formulated the problem of emergency hub placement in the aftermath of a natural disaster as the task of finding the minimum dominating set of a graph, and addressed it using classical and quantum approaches.

Leveraging the fact that neutral-atom quantum computers can natively sample independent sets, we introduced an algorithm for approximating the mDS problem by sampling mISs and refining them using a greedy approach. In our method, the generated samples are processed through two branches, one using directly the mISs and the other exploiting complements of large ISs. Both branches are able to find feasible dominating sets of small size, hence a near-optimal approximation of the exact mDS.

We consider both a fully classical pipeline, employing a classical mIS sampler, and a quantum–classical pipeline in which a neutral-atom processor is used as the sampler, with sampled configurations corrected to enforce independence and maximality. We compare the performances of these two approaches, emulating the quantum methods on classical hardware. The results on a pool of synthetic instances of variable size (up to $100$ nodes) indicate that the two approaches achieve similar performance overall, with the quantum-enhanced workflow exhibiting a modest advantage for larger problem sizes. In particular, the solutions found by both methods are on average within two nodes of the exact optimum.

Following the benchmark analysis, we performed experiments on Pasqal’s Fresnel neutral-atom QPU using two realistic emergency hub placement instances in northern Sicily and the Po Valley. For the smaller Palermo case ($|V| = 41$), the hybrid approach recovers the exact mDS via the $\overline{\mathrm{IS}}$ branch despite hardware noise. For the larger Modena instance ($|V| = 100$), which fully utilizes the QPU register, noise effects are more pronounced compared to emulation. Nevertheless, the sampled configurations still enable identification of the same solution as in emulation, demonstrating a certain degree of robustness. Although the optimal solution is not achieved in this case, the algorithm produces a near-optimal mDS within just one node of the exact solution.

Overall, our results suggest that neutral-atom processors in analog mode can already support graph optimization problems beyond their original proposal as solvers for the maximum independent set, when integrated within simple classical pipelines. These case studies provide evidence that current neutral-atom technology can be incorporated into workflows for realistic optimization problems, indicating its potential relevance for SAR applications.

As hardware matures, the gap between emulation and QPU execution is expected to shrink, potentially increasing the utility of neutral-atom sampling within hybrid combinatorial optimization pipelines. Additional performance improvements can be achieved through enhanced post-processing refinement of collected samples and systematic optimization of the QPU analog schedule within a variational framework.

\newpage

\clearpage
\bibliography{lit}
\bibliographystyle{quantum}

\end{document}